\documentclass[preprint]{emulateapj}
\usepackage{epsfig} 
\usepackage{epstopdf}
\usepackage{graphicx}
\usepackage{amsmath} 
\usepackage[colorlinks=true, citecolor=blue,urlcolor=blue]{hyperref}
\usepackage{tabularx}
\usepackage{aas_macros}

\usepackage{breakcites}

\newcommand{\HII}{\mbox{H II}}
\newcommand{\OIII}{\mbox{[O III]}}

\newcommand{\NII}{\mbox{[N II]}}
\newcommand{\SII}{\mbox{[S II]}}
\newcommand{\Ha}{H$\alpha$}
\newcommand{\Hb}{H$\beta$}

\newcommand{\NIIHa}{\NII/H$\alpha$}
\newcommand{\SIIHa}{\SII/H$\alpha$}

\newcommand{\OIIIHb}{\OIII/H$\beta$}
\newcolumntype{R}{>{\centering\arraybackslash}X}

\begin{document}

\title[]{The Role of Radiation Pressure in the Narrow Line Regions of Seyfert Host Galaxies}

\author{Rebecca L. Davies\altaffilmark{1}, Michael A. Dopita\altaffilmark{1,2}, Lisa Kewley\altaffilmark{1,3}, Brent Groves\altaffilmark{1}, Ralph Sutherland\altaffilmark{1}, Elise J. Hampton\altaffilmark{1}, Prajval Shastri\altaffilmark{4}, Preeti Kharb\altaffilmark{4}, Harish Bhatt \altaffilmark{4}, Julia Scharw\"achter\altaffilmark{5}, Chichuan Jin\altaffilmark{6}, Julie Banfield\altaffilmark{1,7},  Ingyin Zaw\altaffilmark{8}, Bethan James\altaffilmark{9}, St\'ephanie Juneau\altaffilmark{10},  \& Shweta Srivastava\altaffilmark{11}}
\email{Rebecca.Davies@anu.edu.au}

\altaffiltext{1}{Research School of Astronomy and Astrophysics, Australian National University, Canberra, ACT 2611, Australia.}
\altaffiltext{2}{Astronomy Department, King Abdulaziz University, P.O. Box 80203, Jeddah, Saudi Arabia.}
\altaffiltext{3}{Institute for Astronomy, University of Hawaii, 2680 Woodlawn Drive, Honolulu, HI 96822, U.S.A.}
\altaffiltext{4}{Indian Institute of Astrophysics, Sarjapur Road, Bengaluru 560034, India}
\altaffiltext{5}{LERMA, Observatoire de Paris, PSL, CNRS, Sorbonne Universit\'es, UPMC, F-75014 Paris, France}
\altaffiltext{6}{Max-Planck-Institut f\"{u}r Extraterrestrische Physik, Giessenbachstrasse, D-85748 Garching, Germany}
\altaffiltext{7}{ARC Centre of Excellence for All-Sky Astrophysics (CAASTRO)}
\altaffiltext{8}{New York University (Abu Dhabi) , 70 Washington Sq. S, New York, NY 10012, USA }
\altaffiltext{9}{Institute of Astronomy, Cambridge University, Madingley Road, Cambridge CB3 0HA, UK }
\altaffiltext{10}{CEA-Saclay, DSM/IRFU/SAp, 91191 Gif-sur-Yvette, France}
\altaffiltext{11}{Astronomy and Astrophysics Division, Physical Research Laboratory, Ahmedabad 380009, India}

\begin{abstract}
We investigate the relative significance of radiation pressure and gas pressure in the extended narrow line regions (ENLRs) of four Seyfert galaxies from the integral field Siding Spring Southern Seyfert Spectroscopic Snapshot Survey (S7). We demonstrate that there exist two distinct types of starburst-AGN mixing curves on standard emission line diagnostic diagrams which reflect the balance between gas pressure and radiation pressure in the ENLR. In two of the galaxies the ENLR is radiation pressure dominated throughout and the ionization parameter remains constant ($\log U \sim 0$). In the other two galaxies radiation pressure is initially important, but gas pressure becomes dominant as the ionization parameter in the ENLR decreases from $\log U \sim 0$ to $-3.4 \la \log U \la -3.2$. Where radiation pressure is dominant, the AGN regulates the density of the interstellar medium on kpc scales and may therefore have a direct impact on star formation activity and/or the incidence of outflows in the host galaxy to scales far beyond the zone of influence of the black hole. We find that both radiation pressure dominated and gas pressure dominated ENLRs are dynamically active with evidence for outflows, indicating that radiation pressure may be an important source of AGN feedback even when it is not dominant over the entire ENLR.
\end{abstract}

\keywords{galaxies: active --- galaxies: Seyfert --- galaxies: ISM}
 
\section{Introduction}
Star formation and accretion onto supermassive black holes are key drivers of galaxy evolution, but the interplay between these two processes remains one of the mysteries of modern astrophysics. Scaling relations between black hole mass and the velocity dispersion, mass and luminosity of the stellar bulge \citep[e.g.][]{Magorrian98, Ferrarese00, Marconi03} suggest that black holes co-evolve with their host galaxies. Stellar ejecta may be an important source of low angular momemtum gas to fuel active galactic nuclei (AGN; \citealt{Davies07, Wild10}) and AGN feedback may regulate star formation in the host galaxy. The complexity of this relationship is emphasised by the range of evolutionary stages in which AGN host galaxies are found, including galaxies on the star formation rate main sequence \citep[e.g.][]{Silverman09, Mullaney12} as well as green valley \citep[e.g.][]{Ka03b, Georgakakis08}, post-starburst \citep[e.g.][]{Cales13} and quiescent galaxies \citep[e.g.][]{Olsen13}.

The impact of AGN activity on star formation is likely to depend strongly on the mode of coupling between the AGN accretion power and the gas in the host galaxy. At low Eddington ratios ($\lambda \la 0.05$) the accretion power couples to the gas mechanically by directly driving relativistic jets which inflate cavities in the hot interstellar medium (ISM) \citep[e.g.][]{Bicknell95}. The energy transferred to the ISM can balance radiative cooling, thereby suppressing further star formation \citep[e.g.][]{Birzan04, Smolcic09}. At the high Eddington ratios characteristic of Seyfert galaxies ($\lambda \ga 0.05$), radiation pressure may drive galaxy scale outflows which have the potential to remove significant amounts of molecular gas from galaxy disks. \citet{Cecil02} resolved high velocity radiation-pressure driven outflows from dense clouds in the narrow line region of NGC~1068. Outflows with velocities of hundreds to thousands of kilometres per second are prevalent in the most luminous AGN \citep[e.g.][]{Trump06, Page11} and may be ubiquitous across the whole AGN population \citep{Ganguly08}. Both ionized gas \citep{Fu09, Harrison12} and molecular gas \citep[e.g.]{Alatalo11} are entrained in the outflows and may eventually be pushed out of the galaxy, inhibiting future star formation and black hole growth. 

The pressure balance in the ENLRs of Seyfert galaxies is an important diagnostic of the degree of coupling between the AGN radiation field and the ENLR gas. Radiation becomes the dominant source contributing to the static gas pressure in a nebula when the ratio of the ionizing photon density to the atomic density (the dimensionless ionization parameter, \mbox{$U = \frac{S_*}{n_Hc}$}) in the ionized gas exceeds \mbox{$\log U$ = -2}. Above this critical value, the AGN NLR optical spectrum becomes almost independent of the initial ionization parameter. The gas density in the low ionization zone (from which the \OIII, \Ha\ and \NII\ lines originate) scales linearly with the ionizing photon flux, and therefore the \emph{local} ionization parameter asymptotes to a constant \citep{Dopita02}. If radiation pressure is dominant on kpc scales, then the luminosity and opening angle of the AGN ionization cone could play a crucial role in determining the rate and velocity of outflows as well as the rate and distribution of star formation within AGN host galaxies \citep[see e.g. discussion in][]{Stern14}.

The commonly used \NIIHa\ and \SIIHa\ vs. \OIIIHb\ diagnostic diagrams \citep{Baldwin81, Veilleux87, Ke01a} are excellent probes of the pressure balance in AGN ENLRs because the positions of spectra on these diagrams are very sensitive to the ionization parameter \citep[especially in the gas pressure dominated regime where $\log U <$ -2;][]{Dopita02, Groves04}. If gas pressure is dominant, then the ionization parameter in the low ionization zone of the ENLR is free to vary and the diagnostic line ratios will also vary within individual ENLRs and between galaxies. If, however, radiation pressure is dominant, then both the local ionization parameter and the diagnostic line ratios will be fixed \citep{Dopita02, Groves04}. Integral field spectroscopy allows us to populate emission line diagnostic diagrams with line ratios extracted from hundreds of spectral pixels (spaxels) within individual galaxies. These diagrams can then be used to probe the variation in the pressure balance and the relative contribution of star formation and AGN activity across AGN host galaxies \citep{Dopita14, Davies14b, Davies14a}.

In this paper, we use integral field spectroscopy data for four galaxies from the S7 survey \citep{Dopita15} to show that both radiation pressure and gas pressure dominated ENLRs exist and that they produce distinct line ratio distributions on the \NIIHa\ and \SIIHa\ vs. \OIIIHb\ diagnostic diagrams. We also show that both radiation pressure dominated and gas pressure dominated ENLRs are dynamically active and display evidence for outflows, suggesting that radiation pressure may be an important source of AGN feedback even when it is not dominant over the entire ENLR. We describe our observations and data processing in Section \ref{sec:obs}, and discuss the line ratio distributions of our four galaxies in Section \ref{sec:mixing_examples}. We describe our model grids and compare them to our observations in Section \ref{sec:models}. We discuss the impact of radiation pressure on the kinematics of the ionized gas in Section \ref{sec:RP} and summarise our conclusions in Section \ref{sec:conclusions}.

\begin{figure*}
\includegraphics[scale=1.05, clip = true, trim = 0 230 0 0]{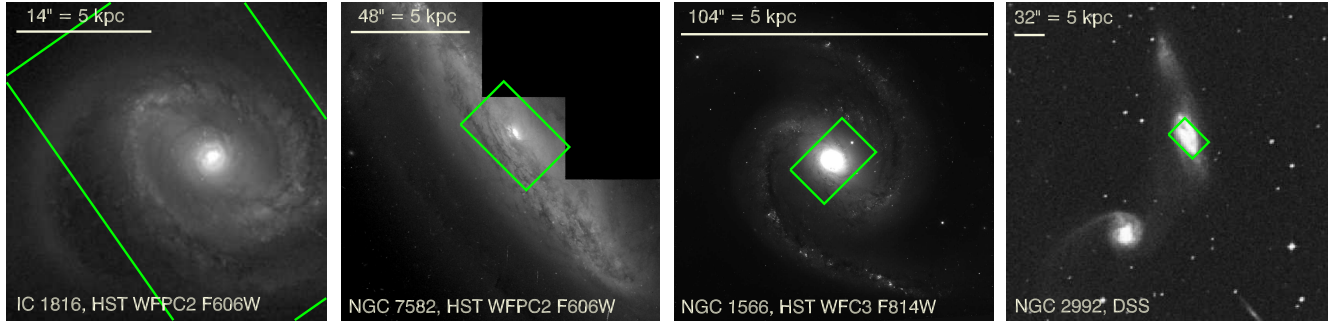}
\caption{Broadband HST and DSS images of the galaxies in our sample with the 38 $\times$ 25 arcsec$^2$ WiFeS FOV overlaid in green. The galaxies have a range of morphologies, inclinations and apparent sizes.} \label{galaxy_mosaic}
\end{figure*}

\section{Observations and Galaxy Sample}
\label{sec:obs}
\subsection{Observations and Data Processing}
S7 is an integral field survey of $\sim$140 low-redshift ($z<0.02$), southern (declination $<10\degr$) AGN host galaxies \citep{Dopita15}. The Wide Field Spectrograph (WiFeS; \citealt{Dopita07,Dopita10}) on the ANU 2.3m telescope is used to observe the central \mbox{$38\times25$ arcsec$^2$} region of each galaxy with a spatial sampling of \mbox{1.0 arcsec pix$^{-1}$}. The seeing of the S7 observations ranges from 1.0-3.0 arcsec (FWHM). The redshift criterion ensures that we achieve a physical resolution of less than 1 kpc for the majority of our targets, which is sufficient to resolve the AGN NLRs. WiFeS is a double-beam spectrograph with three gratings mounted on each arm (one high dispersion grating; R~=~7000 and two low dispersion gratings; R~=~3000). The S7 observations were conducted using the high dispersion red grating (R7000, $\lambda$~=~530-710 nm) and a low dispersion blue grating (B3000, $\lambda$~=~340-570 nm). The high spectral resolution of the red channel observations allows us to isolate multiple kinematic components in the emission line spectra. 

The data were reduced using the Python pipeline \textsc{PyWiFeS} \citep{Childress14} and emission line fitting was performed using the IDL emission line fitting toolkit {\tt LZIFU} (Ho et al., in prep, see \citealt{Ho14} for a brief description of the code).  \textsc{LZIFU} uses \textsc{pPXF} \citep{Cappellari04} to subtract the stellar continuum and then employs the Levenberg-Marquardt least squares fitting routine \textsc{MPFIT} \citep{Markwardt09} to model each emission line spectrum as a linear combination of 1, 2 and 3 Gaussian emission line components. We use an artificial neural network called \textsc{The Machine} (Hampton et al., in prep) to determine the number of Gaussian components required to represent the emission line spectrum of each spaxel. 

\subsection{Galaxy Sample}
In this paper we investigate the balance between radiation pressure and gas pressure across the ENLRs of four S7 galaxies (IC~1816, NGC~7582, NGC~1566 and NGC~2992). These galaxies all have unique distributions of spectra on the \NIIHa\ and \SIIHa\ vs. \OIIIHb\ diagnostic diagrams, and together are representative of all the diagnostic line ratio distributions observed across the S7 sample. 

Figure \ref{galaxy_mosaic} shows broad band \emph{Hubble Space Telescope} (HST) and Digitised Sky Survey (DSS) images of the four galaxies, with the WiFeS FOV overlaid in green. The galaxies have a range of morphologies, inclinations and apparent sizes. IC~1816 is a tidally disturbed Sab galaxy with a bar and a star-forming ring \citep{Malkan98}, and hosts a Seyfert 2 nucleus. The galaxy is virtually face-on (i = 15.2$^\circ$\footnote{Inclination values sourced from the \textsc{HYPERLEDA} database; \href{http://leda.univ-lyon1.fr/}{http://leda.univ-lyon1.fr/}; \citealt{Makarov14}}) and our WiFeS observations cover the central 7.9~$\times$~13.5~kpc$^2$ region at a resolution of 960~pc (FWHM). NGC~7582 is also a Sab galaxy with a clear star-forming disk \citep{CidFernandes01}. The Seyfert 2 nucleus powers a bright extended ionization cone \citep{Morris85}. The galaxy is significantly inclined \mbox{(i = 68.2$^\circ$)} and our WiFeS observations cover the central 2.6~$\times$~3.9~kpc$^2$ region at a resolution of 130~pc. NGC~1566 is a Sb galaxy with a bar and star formation occurring in clear spiral arms \citep[see e.g.][]{Smajic15}, and hosts a Seyfert 1 nucleus. The galaxy has an intermediate inclination to the line of sight \mbox{(i = 47.9$^\circ$)} and our WiFeS observations cover the central 1.2~$\times$~1.8~kpc$^2$ region at a resolution of 60~pc. NGC~2992 is an interacting Sa galaxy with a large biconical ENLR powered by the Seyfert 2 nucleus \citep{Allen99}. NGC~2992 experienced a burst of star-formation 40-50 Myr ago but there is no evidence for current star-formation \citep{Friedrich10}. The galaxy is highly inclined \mbox{(i = 90.0$^\circ$)} and our WiFeS observations cover the central 3.8~$\times$~5.8~kpc$^2$ region at a resolution of 300~pc. 

\begin{figure}
\begin{centering}
\includegraphics[scale=0.6]{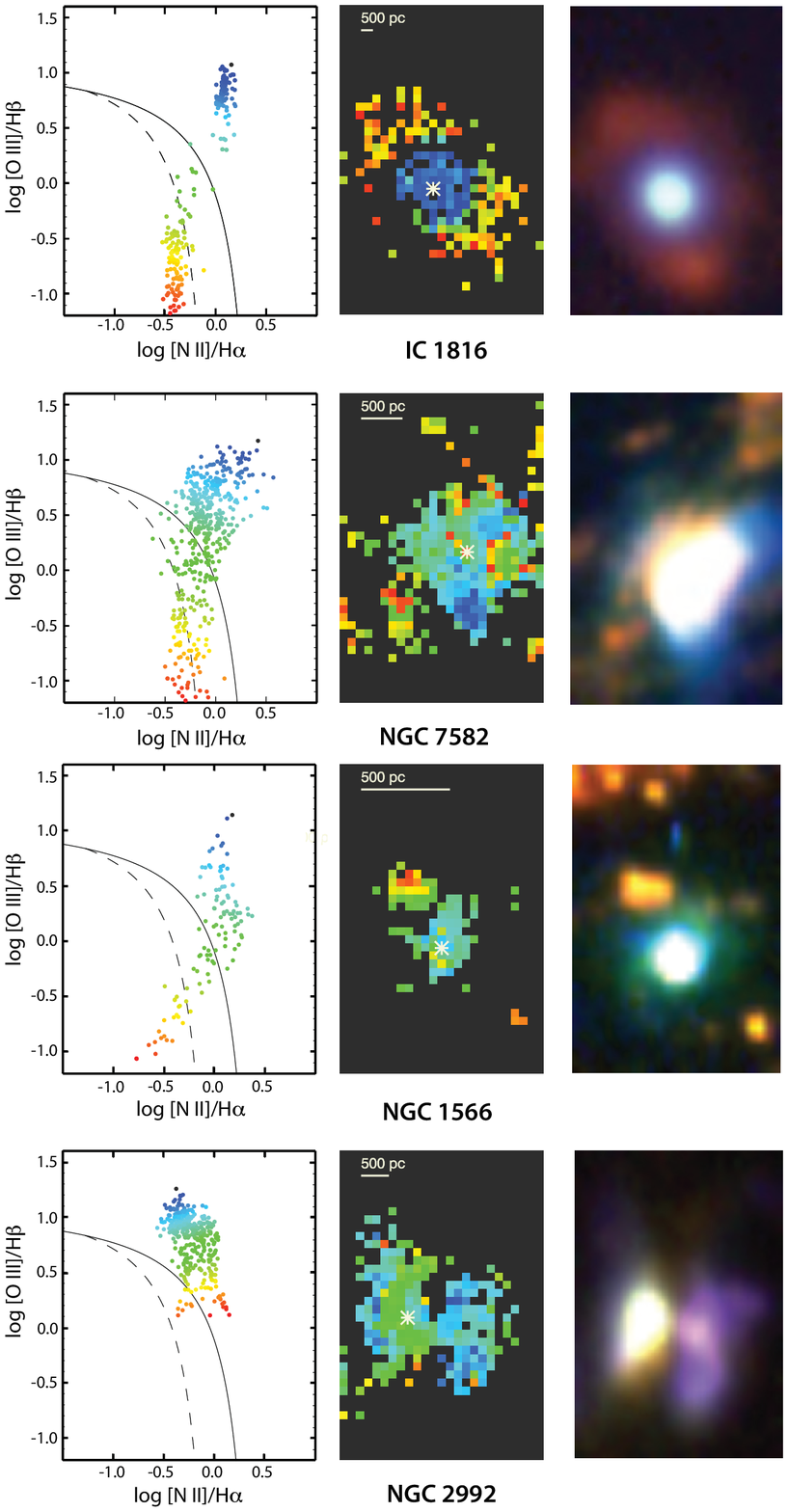}
\end{centering}
\caption{Left: \NIIHa\ vs. \OIIIHb\ diagnostic diagrams for our four representative S7 galaxies, where each datapoint represents the line ratios of an individual velocity component within an individual spaxel. The solid and dashed lines are the \citet{Ke01a} and \citet{Ka03} classification lines respectively. The spectra are color coded according to their positions along the distributions. Centre: Maps of the galaxies using the color coding assigned in the right hand panels. Right: Artificial 3-color images constructed using $H\alpha$ (red), [\ion{N}{2}] (green) and [\ion{O}{3}] (blue). The line ratios of IC~1816 and NGC~7582 vary smoothly between the star forming and AGN dominated regions of the diagnostic diagram, suggestive of mixing between a constant \HII\ region spectrum and a constant ENLR spectrum. NGC~1566 displays a curved mixing sequence which is likely to be indicative of ionization parameter variations within the ENLR. NGC~2992 shows a similar distribution of spaxels in the AGN dominated region of the diagnostic diagram to NGC~1566 but does not show any evidence for current star formation.} \label{fig:mixing_diagrams}
\end{figure}

\section{Mixing Sequences}
\label{sec:mixing_examples}
\subsection{\NIIHa\ and \SIIHa\ vs. \OIIIHb\ diagnostic diagrams}
The \NIIHa\ and \SIIHa\ vs. \OIIIHb\ diagnostic diagrams probe the pressure balance in the AGN ENLR \citep{Dopita02, Groves04} as well as the relative contributions of star formation and AGN activity to observed emission line spectra \citep{Baldwin81, Veilleux87, Ke06}. Spectra dominated by star formation fall along the \HII\ region sequence which traces variations in the ionized gas abundance \citep{Dopita86, Dopita00}. Spectra with some contribution from an AGN lie along the AGN branch which spans from the high metallicity end of the star-forming sequence towards larger \NIIHa, \SIIHa\ and \OIIIHb\ ratios. The hard AGN ionizing radiation field increases the rate of collisional excitation in the nebula and therefore enhances the \NIIHa, \SIIHa\ and \OIIIHb\ ratios. The greater the relative contribution of the AGN, the greater the enhancement of the diagnostic line ratios and the further away from the star forming sequence a spectrum will lie. For this reason, the AGN branch of the diagnostic diagram is often referred to as a `starburst-AGN mixing sequence'.

The left panels of Figure \ref{fig:mixing_diagrams} show the \NIIHa\ vs. \OIIIHb\ diagnostic diagrams for IC~1816, NGC~7582, NGC~1566 and NGC~2992. Each datapoint represents the line ratios extracted from the spectrum of an individual velocity component within an individual spaxel. We require all diagnostic emission lines to be detected at the 3$\sigma$ level to ensure accurate placement of spectra on the diagram. It is not necessary to correct the line ratios for reddening because the lines forming each ratio lie close in wavelength. The solid and dashed curves on the diagnostic diagrams delineate the \citet{Ke01a} theoretical upper bound to pure star formation (Ke01 line) and the \citet{Ka03} empirical classification line (Ka03 line), respectively. Spectra lying above the Ke01 line are dominated by AGN activity, spectra lying below the Ka03 line are dominated by star formation and spectra lying in between the Ke01 and Ka03 lines have significant contributions from both ionization mechanisms.

We observe line ratios spanning from the AGN region all the way to the star forming region of the diagnostic diagrams for IC~1816, NGC~7582 and NGC~1566, consistent with the identification of both AGN activity and current star formation in these galaxies. Variations in the relative contributions of star formation and AGN activity are likely to be important drivers of line ratio variations within these galaxies. In contrast, NGC~2992 does not have any spaxels lying in the star forming region of the diagnostic diagram and has very few spaxels lying in the composite region of the diagnostic diagram, reflecting the fact that young stars are not a significant source of ionizing radiation in this galaxy. 

We further investigate the primary drivers of line ratio variations across the four S7 galaxies by analysing the spatial variations in the line ratios. We color-code the spectra according to their positions along the line ratio distributions (from blue at the largest \NIIHa\ and \OIIIHb\ ratios to red at the smallest \NIIHa\ and \OIIIHb\ ratios), and use these colors to map the variations in line ratios across the galaxies (middle panels). When multiple kinematic components are detected within a single spaxel, the color in the map reflects the line ratios of the component with the greatest \Ha\ luminosity. White asterisks indicate the galaxy centres. We also show artificial 3-color images of the galaxies constructed from their H$\alpha$ (red), \NII\ (green) and \OIII\ (blue) emission for comparison (right panels). In these 3-color images, star formation dominated spaxels appear red or orange whereas spaxels dominated by the AGN ENLR appear blue, purple or green.

We observe strong radial trends in the emission line ratios of IC~1816 and NGC~1566. Emission from \HII\ regions is traced by spaxels colored red or orange in the diagnostic diagrams. These spaxels have the smallest line ratios and lie at the largest radii where the contribution from the AGN is expected to be the smallest. Turquoise and blue spaxels arise from the AGN-dominated ENLR in the central regions of the galaxies, and green spaxels are located at intermediate radii with contributions from both star formation and AGN activity. The clear radial variation in the emission line ratios is strongly indicative of mixing between star formation and AGN activity \citep[see e.g.][]{Davies14b,Davies14a}. The relative contribution of the AGN to the observed spectra is greatest in the nuclear regions of these galaxies and decreases with increasing galactocentric distance. 

We also observe mixing between star formation and AGN activity across the nuclear region of NGC~7582, although the largest line ratios are offset from the nucleus and are associated with an asymmetric ionization cone which is clearly visible in the 3-color image. The large dispersion in the mixing sequence of NGC~7582 is likely to be attributable to metallicity and/or ionization parameter variations within this ionization cone (see Section \ref{subsec:model_results}). NGC~2992 similarly displays a prominent asymmetric ionization cone from which the largest line ratios arise. There is no evidence for current star formation in this galaxy, so the progression from blue to red spaxels traces a decrease in the ionization parameter rather than a decrease in the relative contribution of the AGN ionizing radiation field to the observed spectra (see Figure \ref{fig:models} and Section \ref{subsec:model_results}).

\subsection{Classical and Hybrid Mixing Sequences}
We use the line ratio distributions presented in Figure \ref{fig:mixing_diagrams} to infer the dominant source contributing to the static gas pressure in the ENLRs of IC~1816, NGC~7582, NGC~1566 and NGC~2992. The line ratio distributions fall into two clear categories. Suppose we define the line ratios of the intrinsic AGN NLR spectrum to be the line ratios of the spectrum at the uppermost point of the mixing sequence of each galaxy (with the largest \OIIIHb\ ratio) and the line ratios of the intrinsic \HII\ region spectrum to be \mbox{$\log$ (\NIIHa) $\sim -0.4$}, \mbox{$\log$ (\OIIIHb) $\sim -0.7$}. All intermediate spectra on the diagnostic diagrams of IC~1816 and NGC~7582 fall on or close to a mixing line defined by a linear combination of these two spectra. (The spectra with \mbox{$\log$(\NIIHa)$<$ -0.4} are consistent with pure star formation but at higher metallicities; \citealt{Dopita86, Dopita00}). The line ratio distributions of IC~1816 and NGC~7582 are therefore prototypical examples of what we term `classical mixing sequences'. The \emph{local} ENLR spectrum (and therefore the ionization parameter) must be nearly invariant with distance from the AGN. Other examples of this type of mixing have been found in NGC~5427 \citep{Dopita14}, NGC~7130 \citep{Davies14a} and in a number of galaxies drawn from the CALIFA sample \citep{Davies14b}. The lack of variation in the ionization parameter within the ENLRs of classical mixing galaxies suggests that the AGN radiation field is the dominant source contributing to the static gas pressure in the nebula \citep{Dopita02, Groves04}. 

The line ratios of the spectra extracted from the datacube of NGC~1566 are not distributed along a single mixing curve, but delineate a ``hybrid mixing'' curve in which the line ratios first follow an AGN dominated locus and then follow a starburst-AGN mixing locus to the \HII\ region dominated sequence. This type of mixing curve was first reported by \citet{Scharwaechter11}. The presence of an AGN dominated locus (rather than a single AGN spectrum as observed in the classical mixing cases) indicates that there must be spatial variations in the local ENLR spectra across NGC~1566. The line ratios of NGC~2992 do not show any evidence for mixing between star formation and AGN activity, but follow the same AGN-dominated locus as NGC~1566 (see Figure \ref{fig:models}) with increased dispersion driven by excitation variations within the ENLR (similar to those observed in NGC~7582). The observed variations in the local ENLR spectra of NGC~1566 and NGC~2992 indicate that radiation pressure cannot be dominant across their ENLRs. In Section \ref{subsec:model_results} we show that the line ratio variations along the AGN dominated locus of the hybrid mixing galaxies are consistent with ionization parameter variations within their ENLRs.

\newpage

\section{Models}
\label{sec:models}

We compare the observed mixing sequences to line ratios extracted from grids of \HII\ region and AGN NLR photoionization models to infer the range of ionization parameters associated with the ENLR spectra in the classical and hybrid mixing cases. We show that the line ratio variations along the AGN dominated locus of the hybrid mixing sequence are consistent with ionization parameter variations in the ENLR gas, suggesting that gas pressure is dominant. On the other hand, the AGN ENLR spectra of the classical mixing sequences require a constant and very high ionization parameter, as would be expected if radiation pressure is dominant.

\subsection{Description of Models}
The {\tt Mappings 5.0} code{\footnote{Available at \href{miocene.anu.edu.au/Mappings}{miocene.anu.edu.au/Mappings}} (Sutherland et al., in prep.) has been used to construct a grid of AGN NLR photoionization models and a grid of \HII\ region photoionization models. This code is the latest version of {\tt Mappings 4.0} \citep{Dopita13}, with numerous upgrades to both the input atomic physics and the methods of solution.

The diagnostic optical emission line ratios of \HII\ region and NLR spectra are determined by the ionization parameter ($\log U$) and the chemical abundance set \citep[see e.g.][]{Dopita13, Groves04}. We obtain the abundances of the main coolants (H, He, C, N, O, Ne, Mg, Si and Fe) from the local galactic concordance (LGC) abundances of \citet{Nieva12}, based on early B-star data in the solar neighbourhood. Light element abundances are obtained from \citet{Lodders09} and abundances for all other elements are obtained from \citet{Scott15a, Scott15b} and \citet{Grevesse10}. The grid is therefore not strictly based on the solar abundance set, but for convenience we refer to the LGC abundance set as``solar". The full description of the chemical abundance set is given in Nicholls et al. (submitted).

The input \HII\ region EUV spectra are derived from the {\tt Starburst 99} code \citep{Leitherer99} as described in \citet{Dopita13}. The input AGN spectra are derived from the Jin \& Done three-component models \citep{Done12, Jin12b, Jin12c, Jin12a} as described for Model A in \citet{Dopita14}. We adopt a characteristic AGN black hole mass of $10^7$ M$_{\odot}$, and a Seyfert-like Eddington ratio of 0.1, which results in a bolometric luminosity of $\log ({\mathrm L_{bol}/L_\odot}) = 44.2$. The scale factors of the hard and soft Compton components are set to 0.2 and 0.5, respectively. 

\subsection{Model Grids}\label{grids}
Our grids are computed with chemical abundance sets of 3.0, 2.0, 1.5, 1.0, 0.7, 0.5, 0.3, 0.2 and 0.1 times the ``solar'' value defined in the previous section. All photoionization models, whether of \HII\ regions or AGN NLRs, are isobaric with an initial pressure of $\log {\mathrm P/k} = 6.5$~cm$^{-3}$ K (corresponding to an initial density of \mbox{$n = 10^{2.5} \, \rm cm^{-3}$} at \mbox{$T = 10^4$ K}). Since these are non-dynamical models in pressure equilibrium, the local gas pressure in the models increases as the pressure in the radiation field is transferred to gas pressure through photon absorption \citep{Dopita02}. Heating due to photoelectric charging of grains is accounted for in the models \citep{Groves04}. The models are radiation bounded and are stopped when 99 per cent of the hydrogen gas has recombined.

The \HII\ region models span the ionization parameter range \mbox{$-3.5 \leq \log U \leq -2.0$} with $\Delta (\log U) = 0.25$. The ionization parameter $U$ is calculated at the inner edge of the photoionized plasma, closest to the ionizing source. For the AGN NLR models we cover a much wider range in ionization parameter \mbox{($-3.5 \leq \log U \leq 0$)} which accounts for the potential variation in the local ionizing radiation field across the zone of influence of the active galaxy. 

At high ionization parameters ($\log U \geq -1.5$), Compton heating in the leading edge of the photoionized plasma becomes important, and the electron temperature rises to $10^5$K or even higher. The clearest indicator of the presence of such highly ionized gas is coronal emission lines. Coronal lines are often weak but are found in the nuclear spectra of many of the S7 targets \citep{Dopita15}, indicative of very high ionization parameters close to the central AGN. We do not accurately model the coronal line region and therefore we are unable to compare the observed and predicted coronal line strengths. The diffuse ionizing spectrum produced in the Compton heated zone is re-absorbed by the cooler plasma at larger radii.

\begin{figure}[htb!]
\begin{centering}
\includegraphics[scale=0.61]{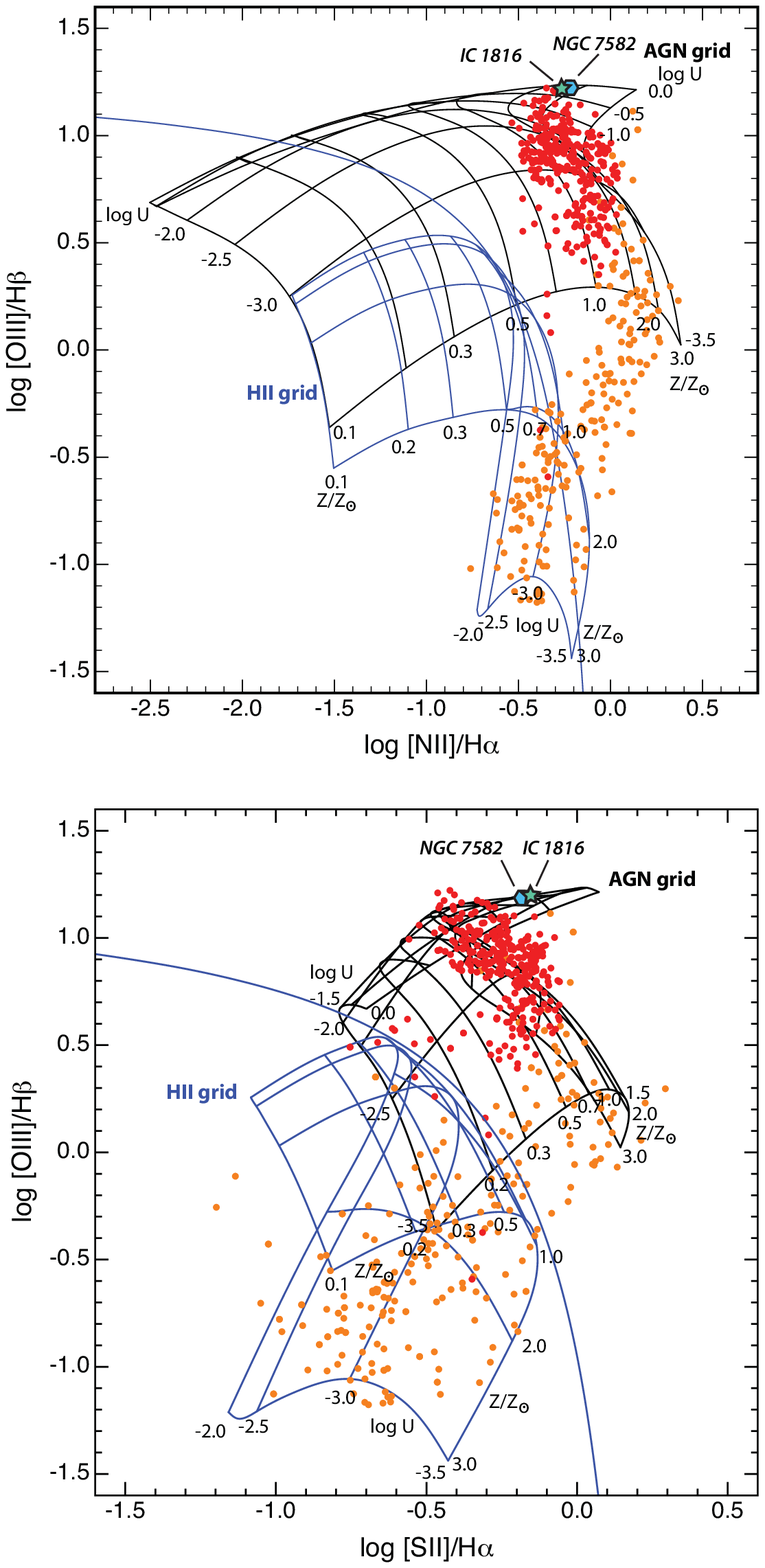}
\end{centering}
\caption{Theoretical grids for AGN NLRs (black) and \HII\ regions (blue) on the \NIIHa\ and \SIIHa\ vs. \OIIIHb\ diagnostic diagrams. The line ratio distributions of NGC~2992 (red) and NGC~1566 (orange) and the upper extremes of the mixing sequences of IC~1816 (green star) and NGC~7582 (blue hexagon) are shown for comparison. The AGN ENLR spectra of NGC~1566 and NGC~2992 require a wide range in ionization parameters ($-3.2 < \log U < 0$). Once the ionization parameter falls below $\sim -3.2 {\rm~to} -3.4$, emission from \HII\ regions becomes important and a standard mixing sequence between the \HII\ region and AGN ENLR spectra is observed. The AGN ENLR spectra of IC~1816 and NGC~7582 require a very high ionization parameter of $\log U \sim$0.0, consistent with them being radiation pressure dominated.} \label{fig:models}
\end{figure}

\subsection{Comparison to Observations}
\label{subsec:model_results}
Figure \ref{fig:models} shows the distribution of the AGN (black) and \HII\ region (blue) grids on the \NIIHa\ vs. \OIIIHb\ (top) and \SIIHa\ vs. \OIIIHb\ (bottom) diagnostic diagrams. We also show the line ratio distributions of NGC~1566 (orange) and NGC~2992 (red) and the upper extremes of the mixing sequences of IC~1816 (green star) and NGC~7582 (blue hexagon) for comparison. 

The AGN dominated spectra in IC~1816 and NGC~7582 are concentrated around $\log$(\OIIIHb) $\sim$ 1.2, requiring an extremely high ionization parameter ($\log U\gtrsim 0.0$). Coronal line emission has been observed in the nuclear regions of both IC~1816 \citep{Dopita15} and NGC~7582 \citep{Reunanen03}, providing independent confirmation of the presence of highly ionized gas. The presence of highly ionized gas across the ENLRs of IC~1816 and NGC~7582 is consistent with the hypothesis that the ENLRs of classical mixing galaxies remain radiation-pressure dominated as they become increasingly contaminated with emission from star forming regions. We note that metallicity variations are expected to induce spread in the \NIIHa\ ratios at a given \OIIIHb\ ratio, and therefore the dispersion in the mixing sequence of NGC~7582 may reflect metallicity variations within the ionization cone. 

The two hybrid mixing galaxies have very similar line ratio distributions to one another in the AGN region of the diagnostic diagrams. The AGN dominated spectra very closely follow the model predictions for ionization parameters varying from \mbox{$\log U \sim 0$} to \mbox{$\log U \sim -3.2 {\rm~to} -3.4$} at super-solar metallicity (\mbox{Z = 1-3 Z$_\odot$}). The spectra with the largest \OIIIHb\ ratios arise from highly ionized, radiation pressure dominated regions. Coronal line emission has been observed in NGC~2992 \citep{Shuder80, Winkler92, Reunanen03}, confirming that much of the gas in the nuclear regions is highly ionized. No coronal lines have been observed in the nuclear region of NGC~1566, but the nuclear emission is strongly attenuated (\mbox{\Ha/\Hb\ = 7.75}, which corresponds to $A_V \sim 3.70$ assuming the \citealt{Fischera05} extinction curve with R$_V^A$ = 4.5 and an unreddened \Ha/\Hb\ ratio of 2.86) and therefore any coronal emission is likely to be undetectable. Gas pressure quickly becomes dominant moving away from the central AGN, allowing the local ionization parameter to be modulated by the gas density. As the ionization parameter decreases even further ($\log U \lesssim -3.4$), emission from \HII\ regions becomes increasingly dominant. The spaxels which lie between the AGN NLR grid and the \HII\ region grid can be explained as a linear sum of \HII\ region spectra with $\log U \sim -2.5 {\rm~to} -3.2$ and AGN NLR spectra with $\log U \sim -3.2 {\rm~to} -3.4$. Eventually the contribution of the AGN NLR becomes negligible and the line ratios are consistent with pure \HII\ region emission, again at super-solar metallicities. 

We note that the AGN NLR spectra with \mbox{$\log U \sim -3.2 {\rm~to} -3.4$} fall into the `composite' region of the \NIIHa\ vs. \OIIIHb\ diagnostic diagram and lie near the Seyfert-LINER boundary on the \SIIHa\ vs. \OIIIHb\ diagnostic diagram, a region of parameter space commonly associated with shock excitation \citep{Ke06, Rich10, Rich11}. We emphasise that these line ratios can be explained by a consistent AGN NLR photoionization model and do not require shocks (see discussion in \citealt{Allen99}).

\begin{figure*}
\includegraphics[scale=0.95, clip = true, trim = 0 30 0 0]{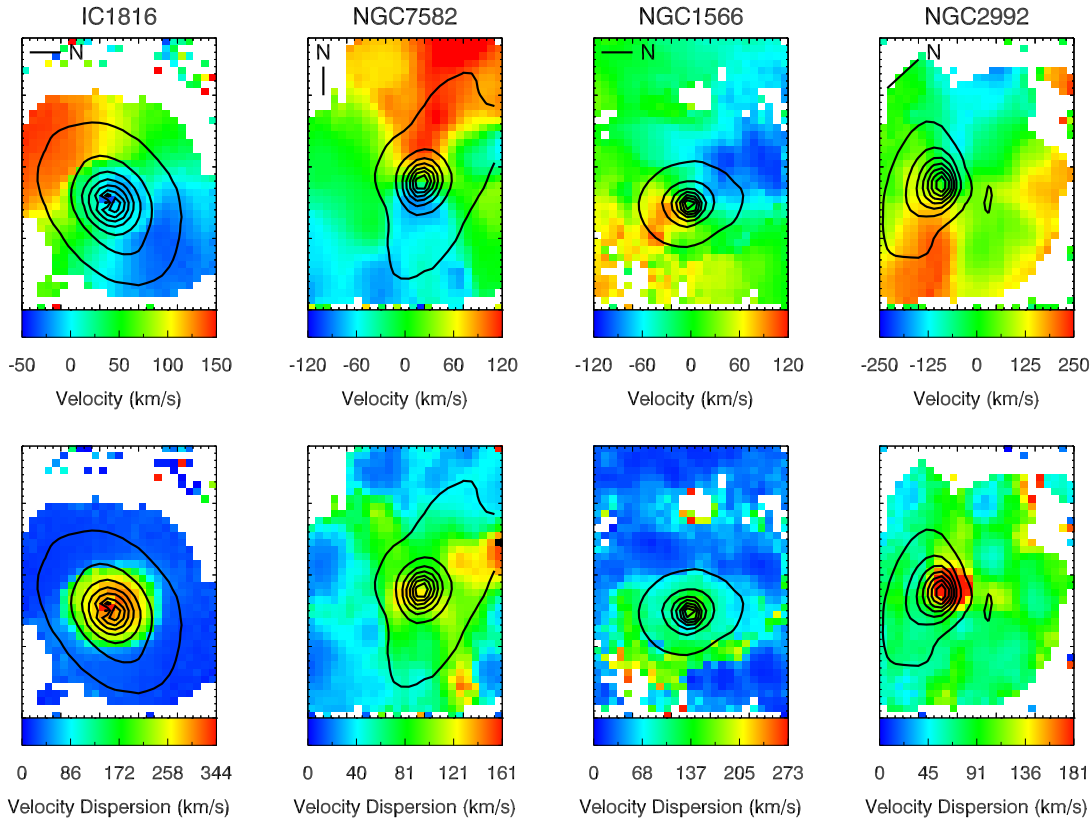}
\caption{Maps of the \Ha\ velocity (top) and velocity dispersion (bottom) fields for the four galaxies in our sample, with contours indicating the spatial distribution of the integrated continuum emission across the galaxies. All four galaxies have dynamically active ENLRs with evidence for outflows.} \label{fig:kinematics}
\end{figure*}

\section{Impact of Radiation Pressure on Gas Kinematics}
\label{sec:RP}
Our investigation of the diagnostic line ratios has revealed that the AGN ionizing radiation field is the dominant source contributing to the static gas pressure across the entire ENLRs of IC~1816 and NGC~7582 but is only dominant in the highest ionization parameter regions of NGC~1566 and NGC~2992. When radiation pressure is dominant it may drive galaxy-scale outflows and therefore be an important source of AGN feedback. An AGN radiation field ionizing a dense cloud in its path will form a strong D-type ionization front at the boundary between the cloud and the photoionized plasma. The radiation field will only drive outflows if optically thin ionized gas is accelerated by the radiation field, which requires the radiation field to be incident on the slab of ionized plasma at an oblique angle. A prototypical example of such an accelerated outflow has been observed in NGC~1068, where comet-tail like filaments are being accelerated away from bright \OIII\--emitting knots at velocities of up to 1500\,kms$^{-1}$ \citep{Cecil02}. In this section we analyse the velocity and velocity dispersion fields of the galaxies in our sample and evaluate the importance of radiation pressure as a source of AGN feedback in radiation pressure dominated and gas pressure dominated ENLRs.

\subsection{Velocity and Velocity Dispersion Fields}
\label{subsec:kinematics}
Figure \ref{fig:kinematics} shows the ionized gas velocity (top) and velocity dispersion (bottom) fields of IC~1816, NGC~7582, NGC~1566 and NGC~2992. The velocities and velocity dispersions are extracted from single component Gaussian fits to the emission line spectra. The black contours indicate the spatial distribution of the integrated continuum emission across the galaxies. All four galaxies show dynamically active ENLRs and have kinematic signatures which are consistent with AGN-driven outflows. 

The star forming disk of IC~1816 displays ordered kinematics consistent with a low velocity dispersion \mbox{($\sigma \sim$30 km s$^{-1}$)} rotation curve. However, the gas in the nuclear region of the galaxy is blue-shifted by \mbox{$\sim$100 km s$^{-1}$} with respect to the underlying rotation curve and has a peak velocity dispersion of \mbox{$\sim$320 km s$^{-1}$}, suggestive of an AGN-driven outflow.

The gas in the plane of NGC~7582 largely follows an ordered rotation curve, but the velocity field of the extra-planar gas is significantly perturbed. The edges of the bipolar ionization cones are delineated by high velocity dispersion \mbox{($\sigma \gtrsim$100 km s$^{-1}$)} regions with significant velocity shifts compared to the underlying rotation curve (\mbox{$\sim$100 km s$^{-1}$}). The line profiles within the ionization cones are often complex and the broad components are blue-shifted with respect to the underlying rotation curve in the south-western cone but red-shifted in the north-eastern cone. The material within the ionization cones is therefore likely being driven away from the galaxy nucleus in an outflow \citep[e.g.][]{Morris85}. NGC~7582 hosts a Compton thick AGN \citep{Turner89, Turner00}, and the radiation pressure is likely to be even stronger in the obscured nuclear region of the galaxy than in the observed circumnuclear region, providing the force required to drive outflows.

The velocity field of NGC~1566 is largely consistent with rotation but shows perturbations due to the presence of multiple kinematic components in the ENLR. The high velocity dispersions \mbox{($\sigma \sim$ 100-200 km s$^{-1}$)} to the south west of the nucleus originate from the superposition of broad blue-shifted emission over the narrow \HII\ region emission which follows the underlying rotation curve of the galaxy. The blue-shifted emission is likely to be associated with outflowing material illuminated by the AGN ionizing radiation field.

The velocity field of NGC~2992 shares many similarities to that of NGC~7582, probably because both galaxies are observed at large inclinations and show prominent ionization cones. The gas in the plane of the galaxy follows a regular rotation curve, but the velocity field of the extra-planar gas deviates significantly from this rotation. Broad and double peaked line profiles are observed in both ionization cones, and the material within the ionization cones is outflowing \citep{Veilleux01}.

The kinematics of the ENLRs indicate that outflows and/or perturbations are ubiquitous among the four galaxies in our sample.

\subsection{Radiation Pressure as a Source of AGN Feedback}
In Section \ref{subsec:kinematics} we showed that both radiation pressure dominated and gas pressure dominated ENLRs are dynamically active and show kinematic signatures of outflows. These results suggest that the presence of radiation pressure dominated regions is sufficient to drive galaxy scale outflows even when gas pressure is dominant over the majority of the ENLR. We note that the observed outflow velocities are unlikely to be sufficient for the outflowing material to escape the galaxies. The observed velocity shifts are only \mbox{$\sim$100-200 km s$^{-1}$} which is comparable to the circular velocities of the galaxies. We conclude that the current levels of AGN activity are unlikely to result in the expulsion of material from these galaxies, but are sufficient to have a significant impact on the density and kinematics of the ISM and may therefore impact star formation activity in the ENLR.

\section{Summary and Conclusions}
\label{sec:conclusions}
We have shown that radiation pressure dominated and gas pressure dominated ENLRs produce distinct line ratio distributions on the \NIIHa\ and \SIIHa\ vs. \OIIIHb\ diagnostic diagrams. The most highly ionized ENLRs are radiation pressure dominated and are associated with classical starburst-AGN mixing sequences. When the ionization parameter exceeds \mbox{$\log U \sim$ -2}, the gas density in the low ionization zone scales with the ionizing photon flux, causing the local ionization parameter (and therefore the AGN spectrum) to be invariant across the ENLR. The ionization parameter remains constant even as emission from \HII\ regions becomes dominant, and therefore the observed line ratios fall along a smooth mixing curve between a single \HII\ region spectrum and a single AGN ENLR spectrum. 

On the other hand, lower ionization parameter ENLRs are dominated by gas pressure and associated with hybrid mixing sequences which include both an AGN dominated locus and a starburst-AGN mixing sequence. When $\log U < -2$, the ionization parameter in the ENLR is moderated by the initial gas density and decreases with increasing galactocentric distance. The \OIIIHb\ ratio decreases with decreasing ionization parameter, producing a locus of AGN dominated spectra on the diagnostic diagrams. At even lower ionization parameters \mbox{($\log U <$ -3.4)}, emission from \HII\ regions becomes important and starburst-AGN mixing is observed. 

We found that radiation pressure may be an important source of AGN feedback regardless of the large scale pressure balance in the ENLR. All four of the galaxies in our sample have dynamically active ENLRs and show kinematic signatures consistent with AGN-driven outflows, indicating that radiation pressure can drive outflows even if it is not dominant across the entire ENLR. The outflow velocities observed for the galaxies in our sample are unlikely to be sufficient for the outflowing material to escape the galaxies. Regardless, the presence of highly ionized, dynamically active regions indicates that radiation pressure is a dominant factor in determining the kinematics and density structure of the ISM in AGN ENLRs, and may therefore have a significant impact on star formation activity. Investigating a larger and more complete sample of galaxies would allow for a more detailed analysis of the impact of the pressure balance on the kinematics of the ENLR, and ultimately would facilitate an investigation of the impact of radiation pressure on star formation activity within AGN ENLRs.

\section{Acknowledgements}
We thank the anonymous referee for their comments which facilitated the improvement of this manuscript. M.D and L.K acknowledge the support of the Australian Research Council (ARC) through Discovery project DP130103925. M.D. would like to thank the Deanship of Scientific Research (DSR), King AbdulAziz University for financial support as Distinguished Visiting Professor under the KAU Hi-Ci program. B.G. gratefully acknowledges the support of the Australian Research Council as the recipient of a Future Fellowship (FT140101202). J.S. acknowledges the European Research Council for the Advanced Grant Program Num 267399-Momentum. JKB acknowledges financial support from the Australian Research Council Centre of Excellence for All-sky Astrophysics (CAASTRO), through project number CE110001020. We acknowledge the usage of the HyperLeda database (\href{http://leda.univ-lyon1.fr}{http://leda.univ-lyon1.fr}). This research has made use of the NASA/IPAC Extragalactic Database (NED) which is operated by the Jet Propulsion Laboratory, California Institute of Technology, under contract with the National Aeronautics and Space Administration.

\bibliography{mybib} 

\end{document}